# High resolution optical time domain reflectometer based on 1.55mm up-conversion photon-counting module

**Matthieu Legré, Rob Thew, Hugo Zbinden, Nicolas Gisin**
*University of Geneva, Group of Applied Physics-Optique*
*20 rue de l'Ecole de médicine, CH-1205 Geneva, Switzerland*
matthieu.legre@physics.unige.ch

*http://www.gap-optique.unige.ch/*

**Abstract:** We implement a photon-counting Optical Time Domain Reflectometer (OTDR) at 1.55µm which exhibits a high 2-point resolution and a high accuracy. It is based on a low temporal-jitter photon-counting module at 1.55µm. This detector is composed of a periodically poled Lithium niobate (PPLN) waveguide, which provides a wavelength conversion from near infrared to visible light, and a low jitter silicon photon-counting detector. With this apparatus, we obtain centimetre resolution over a measurement range of tens of kilometres.

## References and links

## 1. Introduction

Optical Time Domain Reflectometry (OTDR) is the most common distributed measurement technique for fibre network characterization. Photon-counting OTDR (ν-OTDR) is very similar to standard OTDR, except that the detector is a photon-counting module. We have previously shown in [1] that a device, implemented with a Peltier cooled InGaAs photon-counting detector, has an advantage over the standard OTDR technique in providing a better 2-point resolution (~15cm). This resolution is limited by the detection gate duration (~2ns), but the potential improvement of the resolution with this kind of detector is further limited because of its temporal-jitter (400ps). An 850nm ν-OTDR with centimetre-spatial resolution has been implemented with silicon single-photon avalanche diodes (SPAD) in [2]. This is possible because the temporal response of specially designed SPADs can be as small as 40ps, but these detectors are limited to the visible range.

Stimulated by the research of detectors for quantum key distribution, a new class of photon-counting modules has appeared [3-6]. These hybrid devices are based on nonlinear sum frequency generation (SFG) and SPADs. The idea of this kind of photon-counting module is to convert one photon at 1550nm into one visible photon, and then to detect this visible photon with a silicon photon-counting device. A ν-OTDR based on this kind of detection module has already been implemented in [7]. Diamanti *et al.* showed a great reduction of the measurement time, compare to the one needed for a device with a gated detector. Because of their laser pulse duration and the temporal response of their detector, they obtained a 2-point resolution of about 1m. Using detectors, composed of an up-conversion module and a low temporal-jitter SPAD, with very short laser pulses (30ps), we implement a ν-OTDR with a 2-point resolution of 1cm.

## 2. **Photon-counting at 1550nm based on SFG**

The 1550nm photon-counting module we use for this experiment is fully described in [6]. A scheme of the module is depicted in Fig. 1. By SFG, we combine one photon at 1550nm and one at 980nm to obtain one photon at 600nm. The signal (1550nm) and the pump (980nm) are combined with a wavelength division multiplexer (WDM). Then, the two wavelengths are injected with the same polarization into a PPLN waveguide, where the nonlinear process takes place. The up-converted light (600nm) is then spectrally filtered with a prism and interferometric filters before being detected by the SPAD. The SPAD is operating in a passive quenching mode.

As it is explained in [7], a trade-off has to be found between detection efficiency and noise in order to get the best noise-equivalent power (NEP). In our experiment, we have limited the global detection efficiency to 0.8% in order to reduce the noise to 2000Hz, which gives a NEP of $10^{-15}$W/Hz$^{-1/2}$. We use two different SPADs (Micro Photon Device PDM20T, idQuantique ID100) that both exhibit a timing jitter of 40ps. We then have two photon-counting modules at telecom wavelengths with a 40ps timing jitter to implement our ν_OTDR.

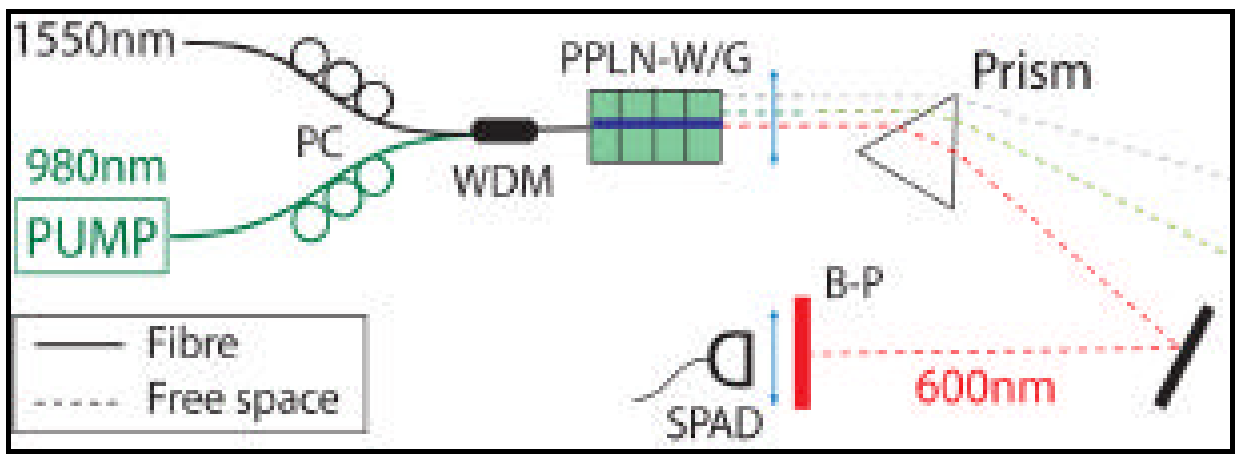

Fig. 1. Set-up of the photon-counting module based on sum frequency generation.

## 3. **Photon-counting OTDR at 1550nm**

The fundamental idea of an OTDR is to measure the time of flight of the light to make a round trip between the device and the point where the light is reflected in the fibre. For this, we use a Time-to-Amplitude Converter (TAC), which gives the time difference between two electrical signals (start and stop). In our experiment, the *start* is trigged by the laser and the *stop* is given by the detector. As we will see, our device can work in two configurations depending on the *start* signal. Depending on the kind of measurement we want to perform, we use either a PicoQuant laser at 1551nm with a temporal width of ~30ps, or a standard DFB telecom laser at 1551nm. The PicoQuant laser is used for the high resolution measurement, whereas the DFB laser is used to obtain long-distance trace. The frequency repetition rate is computed from the length of the system under test. As depicted in Fig. 2, the light pulse is split into two with a 99/1 coupler. When we use an optical triggering (configuration 1), the weak pulse is directly detected by the first detection module (*start*), giving us the reference for the time of flight measurement. In the configuration 2, we use directly the electrical signal used to drive the laser as *start* signal, this signal is delayed compared to the laser trigger with a Digital Delay/Pulse Generator of Stanford Research Systems. After passing through a circulator, the strong pulse is sent into the fibre under test. The back-reflected light is then detected by our second photon-counting module (*stop*). The TAC measures the time difference between the two signals (*start* and *stop*), which is the time of flight of the light for the round trip. The OTDR trace is then obtained by converting the time measured by the TAC into metres. Because our detection module is polarisation dependent, we use a polarisation scrambler before the detector. If this scrambler is switched on, our device is an OTDR. If it is switched off, we have a polarisation-OTDR (P-OTDR), with which we can also study the polarisation properties of optical fibres.

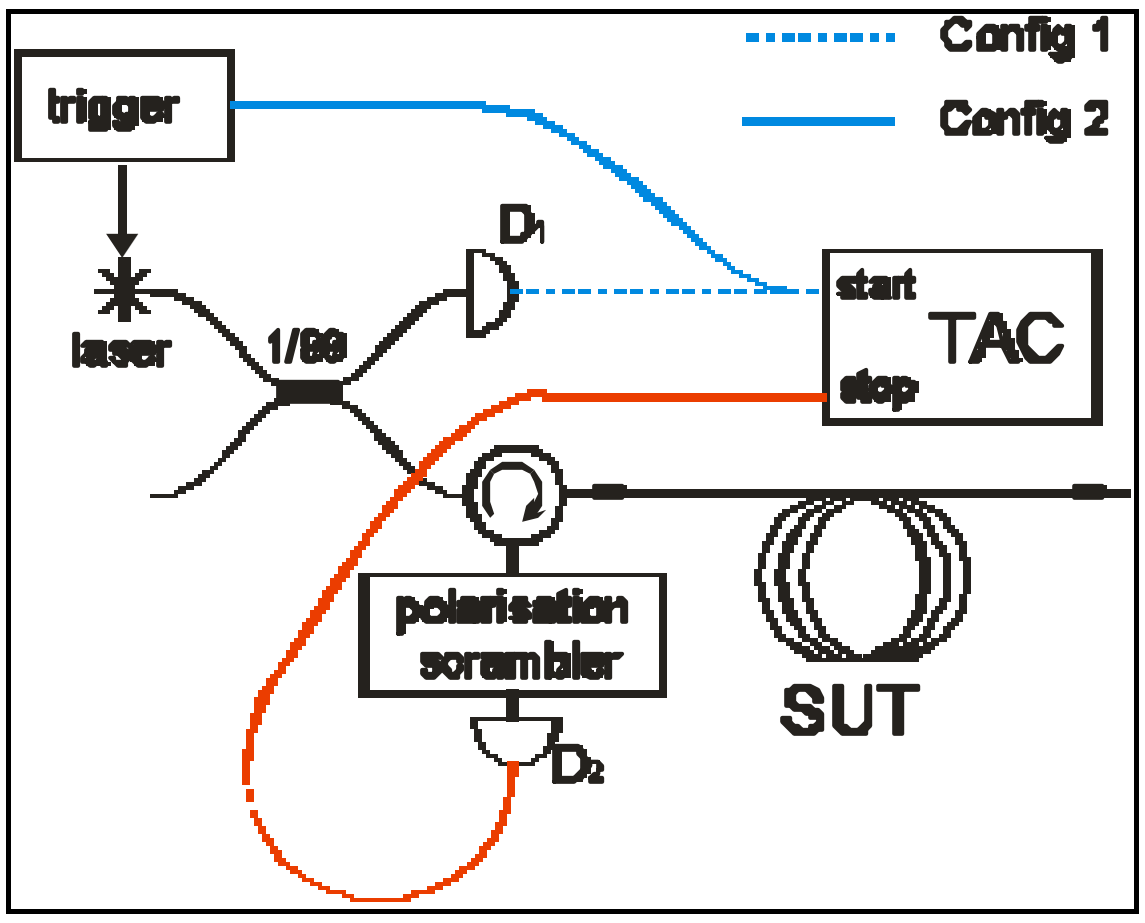

Fig. 2. Set-up of the ν_OTDR. The system is used in either configuration 1 or 2. The system under test (SUT) is composed of two different artefacts either with, or without, a fibre.

An example of OTDR and P-OTDR traces for the same fibre are shown in Fig. 3. The fibre under test has a length of about 16km. In order to see the Rayleigh backscattering signal, we use the following settings: configuration 2; 50ns-width laser pulses; and a TAC resolution of 50ns. With them, we obtain a 2-point resolution of ~7m. Here we focus on standard OTDR-trace characteristics; this is why we use a low resolution. The integration times of the two measurements are not identical, so the absolute values of the two curves can't be compared. On the other hand, we notice that the relative error of the slopes of the best linear fit of the two curves equals 2.5%. The strong fluctuations of the P-OTDR trace correspond to the evolution of the state of polarisation when the light propagates through the fibre. Notice that if we extract the beatlength of the fibre from this measurement, we find the same value as when we compute it from the signal obtained with the ν-P-OTDR we presented in [1]. Considering the settings given above and an integration time of 30 minutes, the peak dynamic range of our device is around 10dB, which corresponds to 50km if we consider an attenuation of 0.2dB/km. If we compute the peak dynamic range as defined in [8], i.e. for a measurement time of 3 minutes, we obtain a value of 5dB. The low efficiency of our detector requires us to integrate measurements over a long time to obtain a good signal to noise ratio. Of course, because the Rayleigh backscattering level depends linearly on the resolution [8], if we improve the resolution of our ν-OTDR, the dynamic of our system is reduced by the same factor.

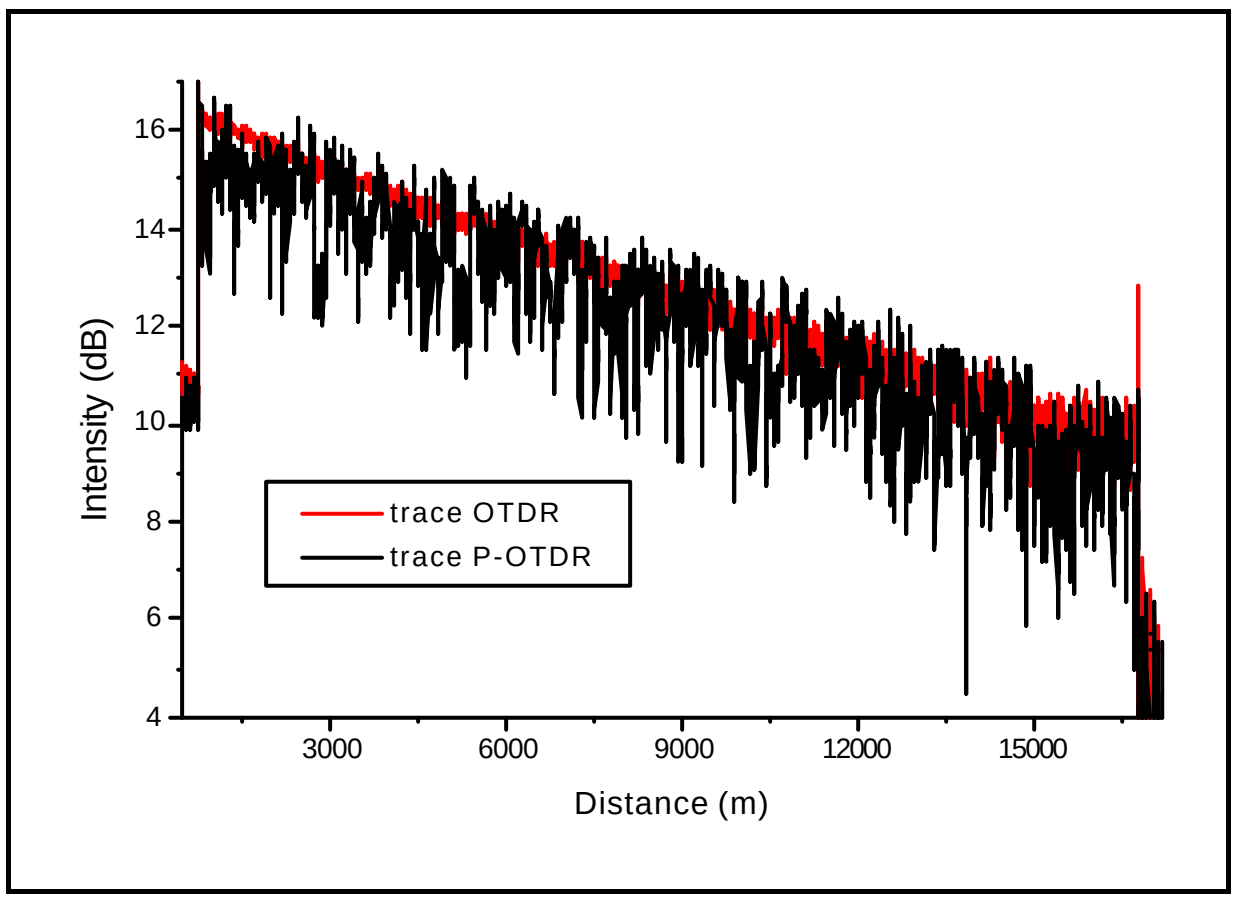

Fig. 3. Measurements of a 16km-length fibre performed by our apparatus. The trace OTDR is obtained when the scrambler is active, the P-OTDR trace whenthe scrambler is off.

## 4. High-resolution and accuracy

In order to evaluate the resolution of our device, we studied different artefacts. The artefact 1 is a 3cm-length U-bench, the artefact 2 is a 4-cm length fibre with a connector at one end and which is cleaved at the other end. The artefact 1 has been used with the configuration 1, and the artefact 2 has been studied with the configuration 2. Notice that the nature of the artefact is not important for the characterisation of the resolution, in both cases we obtained two reflection peaks (in artefact 1, they correspond to the reflections on the two lenses of the U-bench, and in artefact 2, they correspond to the reflections at the connector and the end of the fibre). Let us remind you that the two-point resolution is defined as the minimal distance allowing us to distinguish two reflection peaks. This value is approximately equal to the full-width half-maximum (FWHM) of a single reflection peak [8].

Considering configuration 1, if we zoom in on the U-bench region, as shown by the black curve in Fig. 4-a, we can clearly see two peaks separated by ~3cm of air, corresponding to the two lenses of the U-bench. We notice that in this figure, the width of one peak is around 1.1cm and hence that the resolution is about 1cm. Considering the laser pulse width (30ps) and the temporal jitter of the detectors (40ps), we compute a resolution of ~7mm(=70ps). The difference between the computed value and the measured one can be explained by an additional electronic temporal jitter of the laser and the TAC. We can also notice that we can't see the Rayleigh backscattering, because, as we said above, the dynamic range needed to see Fresnel reflections and the Rayleigh backscattering at the same time is around 70-80dBm for a 2-point resolution of 1cm.

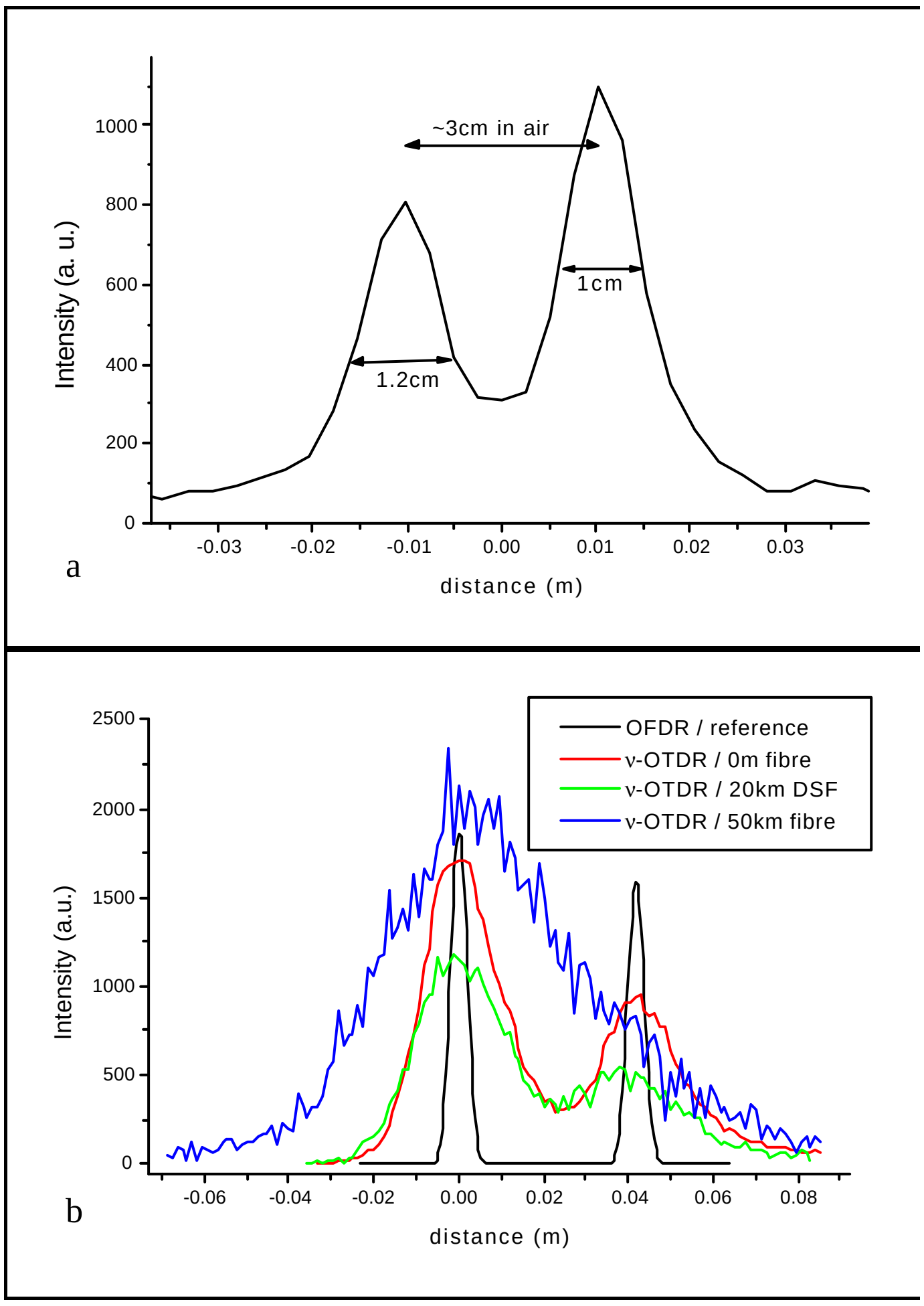

Fig. 4. a_ Measurement of the artefact 1 in configuration 1, the obtained resolution is ~1cm. b_ Measurements of artefact 2 in configuration 2. OFDR: black line; ν-OTDR/0m fibre: red line; ν-OTDR/20km DSF: green line; ν-OTDR/50km fibre: blue line.

To simplify our experiment, we use configuration 2 which allows us to use only one detector instead of two. Unfortunately, as can be seen in Fig. 4-b, this change leads to a slight deterioration of the 2-point resolution (due to a higher jitter level of the start signal). Four curves are drawn in this figure. The black line is the curve obtained with an OFDR, which is another reflectometric device. The OFDR used here has a resolution of around 4-5mm. These devices are generally more precise than OTDRs although they are limited to distances shorter than ~2km [9]. This curve is considered as a reference. The curve measured when a 0m-length fibre is placed before the artefact 2 is drawn with a red line. The measurements performed with a 20km-length of DSF fibre and a 50km-length standard fibre correspond to the green and blue lines respectively. The OFDR measurement gives us a reference length of 4.17cm for the artefact 2. When the artefact 2 is directly measured, our device allows us to resolve the two reflection peaks. The full width half maximum (FWHM) of one peak is 2.1cm, so the 2-point resolution of the apparatus equals ~2cm. If we place a 20km-length dispersion shifted fibre (DSF) before the artefact, we can still distinguish the two peaks of the artefact 2. The FWHM of the reflection peaks remains the same, and hence the 2-point resolution of the device is still ~2cm after 20km of DSF. On the other hand, when a 50km-length standard fibre is put before the artefact, the FWHM of the peaks is strongly increased to ~5.1cm, so in this case, the 2-point resolution is reduced to ~5cm. With a 50km-length fibre, we are not able to distinguish the two peaks anymore. Only the slight asymmetry of the peak let us suppose the presence of the second reflection peak. If we consider that this resolution comes from the combined effects of the response time of the detection scheme (rms=2.0cm) and the laser pulse width (rms=0.3cm), we can compute the theoretical resolution for a Fourrier transform limited pulse [8]. This value equals 5.2cm and the dispersion length [10] equals 6.3km. Theoretical and experimental values agree very well, which means that our device is dispersion limited at 50km. If we consider that the system is limited by the chromatic dispersion when the broadening due to chromatic dispersion is equal to the rms-width coming from the electronic part, then our $\nu$-OTDR is chromatic dispersion limited on measurement range longer than 20km.

The last parameter of our device we have characterized is the accuracy. The accuracy is the precision we have on the measurement of a distance between two peaks. In order to measure it, we use a 2.3m-length pigtail as the system under test. This system is measured with our apparatus ten times and the accuracy is computed as the standard deviation of these ten measurements. We measure a distance of 2.265m±1mm, so our device can reach an accuracy of ~1mm. Notice that we use a resolution-sampling of 1mm, this means that our device is limited by the temporal-jitter of the electronic part of the system (laser driver and detection) and not by the data acquisition. Significant improvements to the accuracy should be achievable by reducing the jitter of the laser driver.

## 5. Conclusion

In conclusion, we have presented the implementation of a $\nu$-OTDR at 1.55$\mu$m with a detector based on the combination of an up-conversion process in a PPLN waveguide and a low temporal jitter silicon photon-counting module. This apparatus can work as an OTDR or a P-OTDR. Its 2-point resolution is of the order of centimetres over ranges as long as 50km. The ultimate resolution, 1cm, is obtained with a configuration using two low temporal jitter photon-counting modules. This configuration allows us to reduce strongly the jitter introduced by the laser driver. In a more convenient configuration, our $\nu$-OTDR has an intrinsic resolution of 2cm, and it is dispersion limited for lengths longer than 20km. For measured distances up to 5m, the accuracy of our device is 1mm.

**Acknowledgements**: Financial support from the Swiss Federal Department for Education and Science (OFES) in the framework of the European COST299 project, from the Swiss NCCR "Quantum photonics", and from EXFO Electro-Optical Engineering is acknowledged.